\documentclass{pasj00}

\begin{document}
\SetRunningHead{K. Watarai}{Analytical Formulae for Optically Thin Accretion Flows}
\Received{2006/10/17}
\Accepted{2006/12/28}

\title{New Analytical Formulae for Optically-Thin Accretion Flows}

\author{Ken-ya \textsc{Watarai} 
  \thanks{Research Fellow of the Japan Society for the Promotion of Science}}
\affil{Astronomical Institute, Osaka Kyoiku University,
Asahigaoka, Kashiwara, Osaka, Japan 582-8582}
\email{watarai@cc.osaka-kyoiku.ac.jp}


\maketitle

\begin{abstract}
In a previous paper, we described new analytic formulae for optically-thick supercritical accretion flows (Watarai 2006, hereafter paper 1). 
Here we present analytic formulae for optically-thin one-temperature accretion flows including the advection-dominated regime,
 using the ``semi-iterative'' method described in paper 1. 
Our analytic formulae have two real solutions. 
The first solution corresponds to the advection-dominated accretion flow (ADAF), 
 and the second solution corresponds to the radiation-dominated accretion flow
 described by Shapiro, Lightman, \& Eardley (the so-called SLE model). 
Both solutions are given by a cubic equation
 for the advection parameter $f$, which is the ratio of the advection cooling rate
 $Q_{\rm adv}$ to the viscous heating rate 
 $Q_{\rm vis}$, i.e., $f=Q_{\rm adv}/Q_{\rm vis}$. 
Most previous studies assume that $f$ is constant ($f \sim 1$ for the ADAF). 
However, it is clear that $f$ should
 be a function of the physical parameters of the radiative-cooling dominated regime. 
We found that the ratio $f$ can be written as a function of the radius,
 mass accretion rate, and viscous parameter $\alpha$. 
Using this formula, we can estimate the transition radius from the inner optically-thin ADAF to the outer optically-thick standard disk, which can be measured using observations of the quiescent state in black hole X-ray binaries. 
\end{abstract}

\section{Introduction}
Optically-thin accretion flows have been widely studied as a likely origin of the power-law 
 spectral component (up to $\sim$ 100 keV) in Galactic black-hole candidates (GBHCs) and 
 nearby low-luminosity active galactic nuclei (LLAGN), including Sgr A*
 in our Galactic Center. 
Thorne \& Price (1975) proposed an optically thin accretion model for the first time, 
 and Shapiro, Lightman, \& Eardley (1976; hereafter SLE) modified and applied the model to X-ray observations of Cygnus X-1. 
Their models were able to explain the observational properties of the low/hard state in GBHCs very well, but the model is thermally unstable;
 i.e. the viscous heating rate cannot balance the radiative cooling rate 
 so that the SLE model is unstable to small-scale perturbations
 (see Piran 1978; Kato et al. 1998).  

The optically thin advection-dominated accretion flow (ADAF) model
 has been investigated extensively since the end of the 1990s
 (Ichimaru 1977; Narayan \& Yi 1994, 1995a, 1995b; Abramowicz et al. 1995). 
This model has been used to interpret the spectra of black hole X-ray binaries
 in their quiescent or low/hard state as an alternative to the SLE solution. 
Since ADAFs have large radial velocities, and the infalling matter carries
 the thermal energy into the black hole,
 advective energy transport can stabilize the thermal instability; 
 thus ADAF models have been widely applied
 to explain observations of Galactic black-hole candidates
 (e.g., Narayan et al. 1996; Hameury et al. 1997), 
 the spectral transition of Cyg X-1 (Esin et al. 1998),
 and the multiwavelength spectral properties of Sgr A*
 (Narayan et al. 1995; Manmoto et al. 1997; Narayan et al. 1997). 
In addition, many ADAF-like models have been proposed including outflows, or convection, etc. 
Since about 2000, these models have been united into one category,
 and began to be called radiatively inefficient accretion flows (RIAFs)
 (Igumenshchev \& Abramowicz 2000; Di Matteo et al. 2000; Abramowicz et al. 2002; Narayan 2002; Igumenshchev et al. 2003; Yuan et al. 2003). 

The self-similar treatment is a powerful tool for investigating the flow properties 
 under various physical conditions
 without complex numerical computations, for example, the optically thin advection-dominated regime (Spruit et a. 1987; Narayan \& Yi 1994, 1995), 
 convection-dominated flows (Narayan et al. 2000; Abramowicz et al. 2002),
 the optically thick advection-dominated regime
 (Wang \& Zhou 1999; Watarai \& Fukue 1999; Fukue 2000),
 self-gravity (Mineshige \& Umemura 1997), 
 the inclusion of magnetic fields (Shadmehri \& Khajenabi 2005; Akizuki \& Fukue 2005), etc. 
In particular, the self-similar solution (hereafter SSS) for the ADAF shows a good agreement with numerical solutions except at the inner and outer boundaries (Narayan et al. 1997). 
Thus, the self-similar treatment seems to be useful as 0-th order approximations. 

To increase the level of approximation in the analytical formulae, 
 one more condition is required. 
One of the main problems with the self-similar treatment is the energy equation.
The energy equation in the canonical self-similar solution
 does not explicitly include radiative cooling,
 due to the parameterization of $f = Q_{\rm adv}/Q_{\rm vis}$, 
 where $Q_{\rm adv}$ and $Q_{\rm vis}$ are the advective cooling rate and the viscous heating rate, respectively.  
Almost all self-similar ADAF solutions assume that $f$=const, 
 so that the radiative cooling is expressed by $Q_{\rm rad} =(1-f) Q_{\rm vis}$. 
However, $f$ should be a function of radius, and other physical parameters 
 when the radiative cooling is comparable to the advective cooling. 
Moreover if $f$ is close to unity, the effect of radiative cooling does not appear 
 in the canonical self-similar solutions. 
Recently, Watarai (2006, hereafter paper 1) extended the self-similar solution
 to the $f \neq $const. regime by including the effect of radiative cooling via $f$ in the optically thick 
 advection-dominated solutions using a ``semi-iterative'' method, and the derived analytic formulae
 agreed well with numerical solutions over a wide range in radius. 
According to his results, $f$ is given by the solution of a quadratic equation, 
 and is a function of the mass accretion rate and radius. 
This raises the question of whether it is possible to find a variable-$f$ solution
 in the optically-thin case using the same method as  in paper 1. 
If we can obtain the variable $f$ solution without complex numerical procedures, 
 it will facilitate comparison with observations of the low/hard state
 of Galactic black-hole candidates or low-luminosity AGNs. 

In this paper, we derive the analytic form of $f$ in optically-thin accretion flows 
 and discuss the basic properties of the analytic solutions. 
In the next section, we introduce the basic equations and the canonical self-similar solutions. 
Our new analytic treatment is presented in section 3. 
In section 4, we comment on the transition zone in bimodal accretion flows.  
The final section summarizes our results. 

\section{Self-Similar and Analytic Solutions for Optically Thin Accretion Flows}

We introduce the ratio of the advective cooling rate, $Q_{\rm adv}$, 
 to the viscous heating rate, $Q_{\rm vis}$, $f$ ($=Q_{\rm adv}/Q_{\rm vis}$), defined in the same way as in previous papers. 
As for an optically thin ADAF, the time scale of energy transport via Coulomb interaction between ions and electrons is longer than that of accreting matter. 
Because the density of accreting gas is very low. 
Thus, such a flow was called an advection-dominated accretion flow,
 i.e., $f \sim 1$, 
and it becomes a radiatively inefficient throughout the whole of the disk.
In this paper we assume that $f$ is not constant, but variable. 
First, the canonical self-similar solutions are presented in subsection \ref{basic}. 
Second, using the self-similar solutions,
 we improve the parameterization of $f$ via a semi-iterative procedure described in subsection \ref{subsec:grobal}.  

\subsection{Basic Equations}
\label{basic}
In this subsection, we formulate the basic equations, which are the same as those given
 in paper 1 except for the radiative cooling term (see also chapter 8 of Kato et al. 1998). 
We assume a steady axisymmetric accretion flow
 ($\partial /\partial t = \partial / \partial \varphi = 0$) and a gas pressure-dominated regime.  
We solve the following hydrodynamic equations in cylindrical coordinates ($r$, $\varphi$, $z$): 
mass conservation, the momentum equation in the radial direction, angular
momentum conservation, hydrostatic balance, and the energy equation, as follows; 
\begin{eqnarray}
\label{mass}
    \dot{M}
     &=& -2\pi r \Sigma v_{r}, 
\\
\label{r-mom}
     v_r \frac{dv_r}{dr}+ \frac{1}{\Sigma}\frac{d\Pi}{dr}
     &=& r \Omega^2- r \Omega_{\rm K}^2 
         - \frac{\Pi}{\Sigma}\frac{d \ln \Omega_K}{dr},  
\\
\label{ang-mom}
    \dot{M} (\ell-\ell_{\rm in}) 
     &=& -2 \pi r^2 T_{r \varphi}, 
\\
\label{hydro}
   \frac{\Pi}{\Sigma} 
     &=& H^2 \Omega_{\rm K}^2, 
\\
\label{energy}
   Q_{\rm vis}^+ 
     &=& Q_{\rm adv}^- + Q_{\rm rad}^-,  
\end{eqnarray}
where $\dot{M}$, $v_r$, $\Sigma$, $\ell$, $\ell_{\rm in}$, and $H$
 are the mass accretion rate, radial velocity, surface density
 , $\Sigma \approx 2 \rho H$,
 specific angular momentum ($\ell=r^2 \Omega$),
 angular momentum at inner edge of the disk, and the scale height, respectively. 
We adopt the Keplarian angular frequency in a Newtonian potential,
 $\Omega_{\rm K}=(GM/r^3)^{1/2}$. 
The $r$-$\varphi$ component of the viscous stress tensor is expressed by $T_{r
\varphi}$, which is related to the total pressure as $T_{r \varphi}= -\alpha
\Pi$, where $\Pi$ is the height-integrated pressure by $\Pi
\approx 2pH$, 
and $\alpha$ is the viscosity parameter (Shakura \& Sunyaev, 1973). 
The last term in equation (\ref{r-mom})
 represents the correction term for the vertical gravity (Matsumoto et al. 1984). 
We assume we are in the gas pressure-dominated regime; thus the equation of state is 
\begin{equation}
p = \frac{k_{\rm B}}{\mu m_{\rm H}} \rho T,
\end{equation}
 where $k_{\rm B}$, $\mu$, and $m_{\rm H}$ are the Boltzman constant, 
 mean molecular weight, and Hydrogen mass, respectively. 

Viscosity generates thermal energy through the differential rotation of turbulently moving gas, 
 and the viscous heating rate is expressed by 
\begin{equation}
\label{qvis}
Q_{\rm vis}^{+} =  r T_{r \varphi} \frac{d\Omega}{dr} 
 \approx  -T_{r \varphi} \Gamma_{\Omega} \Omega,
\end{equation}
where $\Gamma_{\Omega}$ is the linear approximation form of angular velocity ($\Gamma_{\Omega}=- d \ln\Omega/d \ln r $).
The ``net'' internal energy flux at each radius, i.e., advective cooling, is 
 represented by the following formula:
\begin{equation}
\label{qadv}
Q_{\rm adv}^{-} = - \frac{\dot{M}}{2\pi r} T_0 \frac{ds_0}{dr} 
 \approx \frac{\dot{M}}{2 \pi r^2} \frac{\Pi}{\Sigma} \xi.
\end{equation}
The variables $T_0$ and $s_0$
 represent the temperature and the entropy on the equatorial plane,
 and $\xi$ is a dimensionless quantity, 
\begin{equation}
\label{eq:xi0}
\xi = - \frac{1}{\Gamma_{3}-1}
 \left(\frac{d\ln{p_0}}{d\ln{r}} - \Gamma_1 \frac{d\ln{\rho}_0}{d\ln{r}} \right),
\end{equation}
where $\Gamma_3$ is the ratio of the generalized specific heats (e.g., Chandrasekhar 1967),  
\begin{eqnarray}
\Gamma_1 &=& \beta + \frac{(\gamma -1)(4-3\beta )^2}{\beta + 12(\gamma -1)(1-\beta )}, \\
\Gamma_3 -1 &=& \frac{\Gamma_1 -\beta}{4-3\beta}
\end{eqnarray}
where $\gamma$ is the ratio of specific heats and $\beta$ is the gas-pressure normalized by the total pressure. 

By assuming we are in the gas pressure-dominated regime (i.e., $\beta \to 1$)
 and that $p_0 \propto r^{-5/2}$, and $\rho_0 \propto r^{-3/2}$, from the self-similar solutions,
 then $\Gamma_1$ and $\Gamma_3$ tend to $\gamma$. 
Finally in gas-pressure dominant regime, $\xi$ is given by  
\begin{equation} 
\label{eq:xi}
 \xi = \frac{3}{2} \left( \frac{5/3 -\gamma}{\gamma-1} \right).
\end{equation}
Here the range of $\gamma$ is $1 < \gamma < 5/3$. 
If $\gamma$ is exactly equal to 5/3, then the flow is isentropic,
 and no entropy gradient, i.e., $Q_{\rm adv}^-$ = 0. 
In this paper we set $\gamma = 1.5$; thus $\xi = 0.5$. 

The disk is assumed to be optically thin, gas pressure-dominated,
 with radiation produced only by the free-free process:
\begin{equation}
\label{qrad}
Q_{\rm rad}^- = \varepsilon_{\rm brems} \rho^2 T^{1/2} H. 
\end{equation}
where $\varepsilon_{\rm brems}= 1.24 \times 10^{21}$ is the coefficient of the bremsstrahlung emission.
The thermal synchrotron and its inverse Compton cooling
 are important cooling processes as in shown by several authors (Kusunose \& Takahara 1989; Narayan \& Yi 1995; Manmoto et al. 1997; Mahadevan 1997). 
Such cooling processes are significant near the disk inner region (inside 10 $r_{\rm g}$), but rapidly decreases at disk outer region (roughly $F_{\nu} \propto r^{-2}$).  
Hence we ignore the effect of thermal synchrotron and Comptonization in this paper.

\subsection{Canonical Self-Similar Solutions}

To derive the characteristic radial dependence of the above equations, 
we transform the equations (\ref{r-mom}), ($\ref{ang-mom}$),
 (\ref{qvis}), (\ref{qadv}) as follows:
\begin{equation}
\label{r-mom2}
      \Gamma_{v} v_{r}^2 
    + \left(\Gamma_{\Pi} - \frac{3}{2} \right)\frac{\Pi}{\Sigma} 
    - r^2 (\Omega^2-\Omega_{\rm K}^2) = 0,  
\end{equation}
where $\Gamma_{v}=d \ln v_r/d \ln r $ and $\Gamma_{\Pi}=d \ln\Pi/d \ln r $, which can be determined
 from the self-similar solutions. 
The momentum equation in the azimuthal direction is  
\begin{equation}
\label{ang-mom2}
   \dot{M} \Omega_0 \Omega_{\rm K} B = 2 \pi \alpha  \Pi,
\end{equation}
where $B=(1-\ell_{\rm in}/\ell)$ is the boundary term. 
We also assume $\Omega=\Omega_0 \Omega_{\rm K}$, where $\Omega_0$ is constant. 
The energy balance equation is given by 
\begin{equation}
\label{qvis2}
Q_{\rm adv}^{-} = fQ_{\rm vis}^{+},
\end{equation}
where $f$ is constant in canonical treatments. 

The above equations can be solved analytically, which leads to the following solutions.  
We note that all solutions are normalized by the critical accretion rate,
 $\dot{M}_{\rm crit} = L_{\rm E}/c^2 = 2 \pi c r_{\rm g}/\kappa_{\rm es}$,
 and the Schwarzschild radius, $r_{\rm g}=2GM/c^2$,
 where $L_{\rm E}$, $G$, $M$, $c$ are the Eddington luminosity,
 gravitational constant, black hole mass, and speed of light, respectively.
The normalized mass accretion rate and radius
 are expressed as $\dot{m}=\dot{M}/\dot{M}_{\rm crit}$ and $\hat{r}=r/r_{\rm g}$.  
\begin{eqnarray}
\label{ss-omega}
 \Omega &=& \Omega_0 \Omega_{\rm K}, \\
\label{ss-sigma}
 \Sigma &=& \Sigma_0 f^{-1} \alpha^{-1} \dot{m} \hat{r}^{-1/2}, \\
\label{ss-vr}
 |v_r| &=& v_{r 0} \alpha f \hat{r}^{-1/2}, \\
\label{ss-pi}
 \Pi &=& \Pi_0 \alpha^{-1} \dot{m} \hat{r}^{-3/2}, \\
\label{ss-H}
 H &=& H_0  f^{1/2} \hat{r} r_{\rm g}, \\
\label{ss-T}
 T &=& T_0  f \hat{r}^{-1}, 
\end{eqnarray}
where the coefficient with subscript ``0'' represents the physical quantities: 
\begin{eqnarray}
\label{eq:omega0}
\Omega_0 &=& \left[ 1- \Gamma_{v} \frac{\Gamma_{\Omega}^2}{\xi^2} \alpha^2 f^2 -(\Gamma_{\Pi} -3/2) \frac{B \Gamma_\Omega}{\xi} f \right]^{-1/2}. \\
\Sigma_0 &=& \frac{\sqrt{2}}{\kappa_{\rm es}} \left( \frac{2 \xi}{\Gamma_{\Omega} \Omega_0} \right), \\
v_{r0} &=& \frac{c}{\sqrt{2}} \left( \frac{\Gamma_{\Omega} \Omega_{0}}{2 \xi} \right), \\
\Pi_{0} &=& \frac{B \Omega_0 c^2}{\sqrt{2} \kappa_{\rm es}}, \\
H_0 &=& \sqrt{\frac{B \Gamma_{\Omega} \Omega_{0}^2}{\xi}}, \\
\label{eq:t0}
T_0 &=& \left( \frac{k_{\rm B}}{\mu m_{\rm H}} \right)^{-1}
          \left( \frac{B \Gamma_{\Omega} \Omega_{0}^2}{2 \xi} \right) c^2. 
\end{eqnarray}
From the radial dependence of the self-similar solutions,
 we can determine the values of $\Gamma_{\Pi}$, $\Gamma_v$, $\Gamma_{\Omega}$; 
 i.e., in the present parameter sets these constants are given by $\Gamma_{\Omega}=1.5$, $\xi = 0.5$, $\Gamma_{v}=-0.5$, 
 and $\Gamma_{\Pi}=-1.5$, respectively. 
$\xi$ and $\Omega_0$ in equation (\ref{eq:xi}) are then automatically determined,  
 and finally all the coefficients can also be evaluated using equations (\ref{eq:omega0})-(\ref{eq:t0}). 
 
Using these self-similar formulae, the advective cooling and radiative cooling terms are given by 
\begin{eqnarray}
\label{eq:advcool}
 Q_{\rm adv} &=& Q_{\rm adv}^0 f \dot{m} r_{\rm g}^{-1} \hat{r}^{-3}, \\
\label{eq:radcool}
 Q_{\rm rad} &=& Q_{\rm rad}^0 \alpha^{-2} f^{-2} \dot{m}^2 r_{\rm g}^{-1} \hat{r}^{-5/2},
\end{eqnarray}
where 
\begin{equation}
Q_{\rm adv}^0 = \frac{c^3}{\kappa_{\rm es}}
 \left( \frac{B \Gamma_{\Omega} \Omega_{0}^2}{2} \right), 
\end{equation}
and 
\begin{equation}
\label{eq:radcool0}
Q_{\rm rad}^0 = \frac{\varepsilon_{\rm brem} \sqrt{2} c}{\kappa_{\rm es}^2} 
 \left( \frac{k_{\rm B}}{\mu m_{\rm H}} \right)^{-1/2}
 \left( \frac{2 \xi}{\Gamma_{\Omega} \Omega_{0}} \right)^2.
\end{equation}

The parameter dependences of these solutions are identical to those of the canonical self-similar solutions
 by Narayan \& Yi (1994) or Wang \& Zhou (1999) under the condition in which $f=$constant. 

\subsection{Analytic Solutions under the First Iteration}
\label{subsec:grobal}

In the previous self-similar solutions, the effect of the radiative cooling is
 given by $(1-f) Q_{\rm vis}$. 
However, the energy equation does not explicitly include radiative cooling
 because of the constant $f$ assumption. 
Using the same formulation as in paper 1,
 we can explicitly include the radiative cooling term in $f$. 
From the definition of $f$, 
\begin{equation}
f = \frac{Q_{\rm adv}^-}{Q_{\rm vis}^+} = \frac{Q_{\rm adv}^-}{Q_{\rm adv}^-+Q_{\rm rad}^-}=\frac{1}{1+g}, 
\label{eq:f}
\end{equation}
where $g$ is the ratio between radiative cooling and advective cooling. 
From equation (\ref{eq:advcool})-(\ref{eq:radcool0}), the explicit form of $g$ is  
\begin{eqnarray}
g \equiv \frac{Q_{\rm rad}^-}{Q_{\rm adv}^-} 
= g_0 \frac{\dot{m} \hat{r}^{1/2}}{\alpha^{2} f^{3}}, 
\label{eq:g}
\end{eqnarray}
where $g_0$ is the ratio of $Q_{\rm rad0}$ to $Q_{\rm adv0}$,
 i.e., $g_0=Q_{\rm rad0}/Q_{\rm adv0} \propto \Gamma_{\Omega}^{-3} \Omega_{0}^{-3}$.
In this paper the value of $g_0$ is 
$g_0 \simeq 1.24 \times 10^{-2}$ for the present parameter sets. 
Substituting equation (\ref{eq:g}) into equation (\ref{eq:f}),
 we finally obtain a cubic equation for $f$ in nondimensional form: 
\begin{equation}
 f^3 - f^2 + g_0 
 \frac{\dot{m} \hat{r}^{1/2}}{\alpha^2} = 0. 
\label{eq:3ji}
\end{equation}
In the optically thick case, $f$ is independent of $\alpha$,   
 whereas in the optically thin case, $f$ depends on $\alpha$. 
This arises from the difference in the radiative cooling processes between the optically thin 
 and the optically thick regimes. 
The thermal bremsstrahlung cooling depends on the gas density 
and temperature, so changes in density are implicitly included in
 $f$ via the viscosity parameter $\alpha$. 

Consider a cubic equation, $x^3 + a_1 x^2 +a_2 x + a_3 =0$, with real coefficients, $a_1$, $a_2$, and $a_3$. 
First, we evaluate the determinant of the cubic equation, $D = Q^3 -R^2$,
 as follows: $Q \equiv \frac{a_1^{2} -3 a_2}{9}, R \equiv \frac{2 a_1^3 -9a_1 a_2 + 27 a_3}{54}$. 
The coefficients in equation (\ref{eq:3ji}) are $a_1=-1$, $a_2=0$, and $a_3= g_0 \alpha^{-2} \dot{m} \hat{r}^{1/2}$, respectively.   
When $Q$ and $R$ are real and $D>0$, the cubic equation has three real roots (Press et al. 1986). 
In the present study, $Q=1/9$ and $R=(27 a_3 -2)/54$. 
Here we only consider physically plausible solutions, i.e., $0<f<1$,  
 which results in two real solutions: 
\begin{equation}
f_{\rm ADAF}(\theta) = \frac{1}{3} \left[1-2 \cos{ \left( \frac{\theta+2\pi}{3} \right)} \right]
\label{eq:ADAF}
\end{equation}
\begin{equation}
f_{\rm SLE}(\theta) = \frac{1}{3} \left[1-2 \cos{ \left( \frac{\theta+4\pi}{3} \right)} \right]
\label{eq:SLE}
\end{equation}
where $\theta$ is a function of $\hat{r}$, $\dot{m}$, and $\alpha$. 
Here $f_{\rm ADAF}$ and $f_{\rm SLE}$ correspond to 
 advection-dominated and radiation-dominated solutions. 
The explicit form of $\theta$ is given by 
\begin{equation}
\theta (\hat{r}, \dot{m}, \alpha) = \arccos{\left(\frac{27}{2} g_0 \frac{\dot{m} \hat{r}^{1/2}}{\alpha^{2}} -1 \right)}. 
\end{equation}
If we modify $f$ in equations (\ref{ss-omega})-(\ref{ss-T}) to $f(\theta)$, 
 new sets of analytic solutions will be obtained. 
This method is similar to an iteration procedure: using the SSS as a first approximation,
 we improve the solutions for $f$ via the energy equation, which includes the radiative cooling term. 
We note that our present solutions have been obtained by the first iteration. 
Original SSSs ($f=1$ for ADAF) in equation (\ref{ss-omega})-(\ref{ss-T}) are modified by the first iteration (see figure 1, 2). 
Further iteration can expect the improvement of the solutions
 but the first improvement will be most significant. 
In other words, our solutions are still quite rough approximation but there is an advantage that is possible to describe in the analytical form.

\section{New Analytic Solutions} 
\label{sec:compare}

\begin{figure}
  \begin{center}
    \FigureFile(80mm,80mm){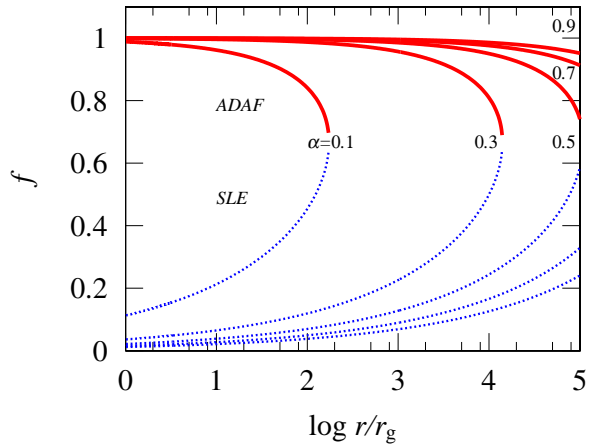}
  \end{center}
  \caption{Radial profiles of $f$ for various values of the viscous parameter $\alpha$.
 Solid and dotted lines represent ADAF-like and SLE-like solutions, respectively. 
 Input parameters are $\alpha=0.1$, 0.3, 0.5, 0.7, and 0.9 from left to right, 
 while the accretion rate $\dot{m}$ is fixed at 0.01. }
\label{fig1}
\end{figure}

We derived the analytic form of $f$ considering the effect of radiative cooling. 
Figure \ref{fig1} shows the solution of $f_{\rm ADAF}(\theta)$ and $f_{\rm SLE}(\theta)$
 for various values of $\alpha$.  
As the values of $\alpha$ increase, the optically thin solutions can extend to the disk's outer regions. 
The $f$ value is close to unity at large radii for high $\alpha$. 
When $\alpha$ is large, the viscosity effectively extracts the angular momentum of the matter; 
 thus advective energy transport will be important over a wide range in radius. 
However, in the outer regions of the disk, radiative losses via bremsstrahlung emission rapidly
 cool the disk as $\alpha$ decreases, so that ADAF-like solutions disappear in the disk's outer region.

\begin{figure}
  \begin{center}
    \FigureFile(80mm,80mm){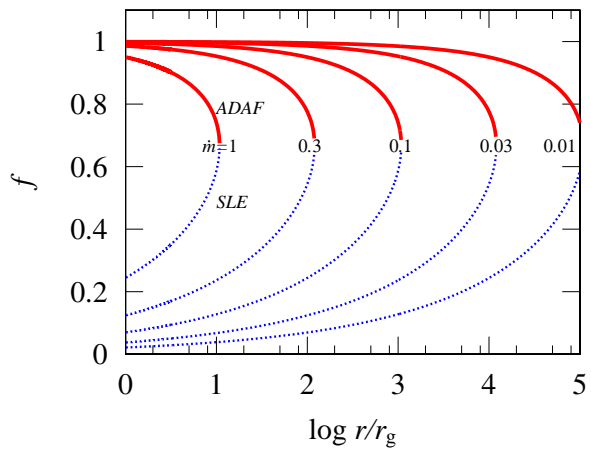}
  \end{center}
  \caption{Same as figure 1, but for $\dot{m}$=0.01, 0.03, 0.1, 0.3, and 1 from right to left. 
The viscosity parameter $\alpha$ is fixed at 0.5. }
\label{fig2}
\end{figure}

Figure \ref{fig2} represents the radial dependence of the mass accretion rate on $f$. 
The optically thin solutions depend on the mass accretion rate as well as $\alpha$. 
The mass accretion rate generally determines the density, 
 which influences the cooling rate via thermal bremsstrahlung, 
 so that if the density is high, the disk will cool more effectively. 
Therefore , the two parameters $\alpha$ and $\dot{m}$ control the basic properties of 
 these solutions through advective and radiative cooling. 

Note that figures 1 and 2 have a logarithmic scale on the horizontal axis
 and a linear scale on the vertical axis. 
Therefore, the optically thin ADAF solutions decrease rapidly at the disk's outer radii. 

Our results do not conflict with the previous $f \sim 1$ result
 for low $\dot{m}$ and high $\alpha$ (Nakamura et al. 1997).
However, according to Nakamura et al. (1997), 
 ADAF solutions for low $\dot{m}$ and small $\alpha$, ($\dot{m}, \alpha$)=($10^{-5}$, $10^{-3}$), 
 and for high $\dot{m}$ and large $\alpha$, ($\dot{m}, \alpha$)=(0.1, 0.1), 
 also show $f \sim 1$ over a wide range of radii. 
This could be due to the outer boundary conditions for the ADAF. 
To discuss the outer region of the optically thin accretion flow,
 it is necessary to account for the outer boundary conditions. 
In some cases outer boundary conditions can affect the disk structure (Yuan 1999; Yuan et al. 2000);
 we comment on the outer boundary conditions in the next section.  

\section{Discussion}
\label{sec:discussion}

\subsection{Transition Radius in Bimodal Accretion Flows} 

Some observations of GBHCs indicate an hot, hard X-ray component and a cold, soft component are coexistent in accretion flow
 (Thorne \& Price 1975, Ichimaru 1977 for Cyg X-1; Shapiro \& Teukolsky 1983;
 Narayan et al. 1996 for soft X-ray transients; 
 Esin et al. 1997 for Nova MUSCA 1991; Narayan et al. 1997 for V404 Cygni). 
To date, two mechanisms have been proposed to explain this hot component: 
 evaporation in the vertical direction of the disk
 (Meyer \& Mayer-Hofmeister 1994) or 
 2-zone radial geometry which the outer parts of the disk follow the optically thick, while the inner disk switches to the optically thin, hot solution, which is secularly stable (SLE). 
 The 2-zone radial geometry in accretion flow was solved self-consistently
 for the first time by Honma (1996),
 who pointed out the importance of thermal conduction for a smooth transition.  
With either mechanism, when radiative cooling works efficiently,
 the hot flow shrinks toward the equatorial plane,
 and the optically thin accretion flow transitions to the standard disk. 
Therefore, the energy transport in the radial direction is important for a smooth transition 
 between the optically thin and thick regimes,
 but it seems that the cooling processes determine the location of the transition radius; this is an important point. 

The location of the transition region between an inner optically thin regime
 and an outer optically thick regime can be evaluated using equation (\ref{eq:ADAF}). 
If we set $f_{\rm ADAF}(\theta)=0.5$ as a critical value of advection-dominated accretion,
 then equation (\ref{eq:ADAF}) has a physical solution ($0<f<1$). 
Then, the transition radius is
\begin{equation}
\label{eq:rtr}
 \hat{r}_{\rm tr} 
\simeq  102 \left(\frac{\alpha}{0.1} \right)^4 \left(\frac{\dot{m}}{0.01}\right)^{-2}
\end{equation}
The parameter dependence of $\hat{r}_{\rm tr}$ is consistent with the results of
 Honma (1996), and is very sensitive to the viscous parameter $\alpha$. 
We have used an independent approach to derive this result. 
Note that equation (\ref{eq:rtr}) represents the radius where
 the solution disappears, but Honma (1996) obtains the real transition radius.

Turbulent thermal conduction in a bimodal accretion flow gives a finite size for the 
 transition zone (Manmoto \& Kato 2000a,b). 
In the transition zone, the heating (cooling) via turbulent thermal conduction is balanced by radiative cooling. 
Turolla \& Dullemond (2000) discussed the flow properties at the evaporation radius in the outflow solution by Blandford \& Begelman (1999); however, they treated the transition radius as an input parameter. 

Recently, Lu et al. (2004) found bimodal solutions for relatively high values of $\alpha$ ($>$0.4)
 without conduction effects. 
They performed numerical integration from the sonic point, 
 and then obtained bimodal solutions. 
However, their method uses a variety of initial conditions at the sonic point 
 so that the determination of the transition radius is difficult. 
Here we note that our procedure for evaluating the transition radius is purely mathematical,
 and the derived transition radius does not depend on the outer boundary conditions. 

Figure 3 represents the radial structure of a bimodal accretion flow based on our model. 
The inner region is described by the optically thin solution,
 and the outer region is a solution with which 
 the standard optically thick, geometrically thin accretion disk
 by Shakura \& Sunyaev (1973) is connected by
 the same accretion rates and $\alpha$ parameters as the inner region.
The inner boundary term in the standard model, $1-\sqrt{r_{\rm in}/r}$,
 was disregarded. Here the $r_{\rm in}$ represents the disk inner radius. 
We ignore radial energy transport between the optically thin regime and the optically thick regime. 
In our solutions, the transition radius is not an input parameter unlike previous studies. 
Our estimate of the transition radius is very sensitive to the angular frequency,
 with $\hat{r}_{\rm tr} \propto g_{0}^{-2} \propto \Omega_{0}^6$, and so we cannot compare our results with previous papers at this stage.  
Our model is very simple, so it will be useful for discussing
 the rough structure of the flow;
 however, numerical calculations will be required for detailed comparisons with observations. 
This will be carried out in future work. 
\begin{figure}
  \begin{center}
    \FigureFile(60mm,60mm){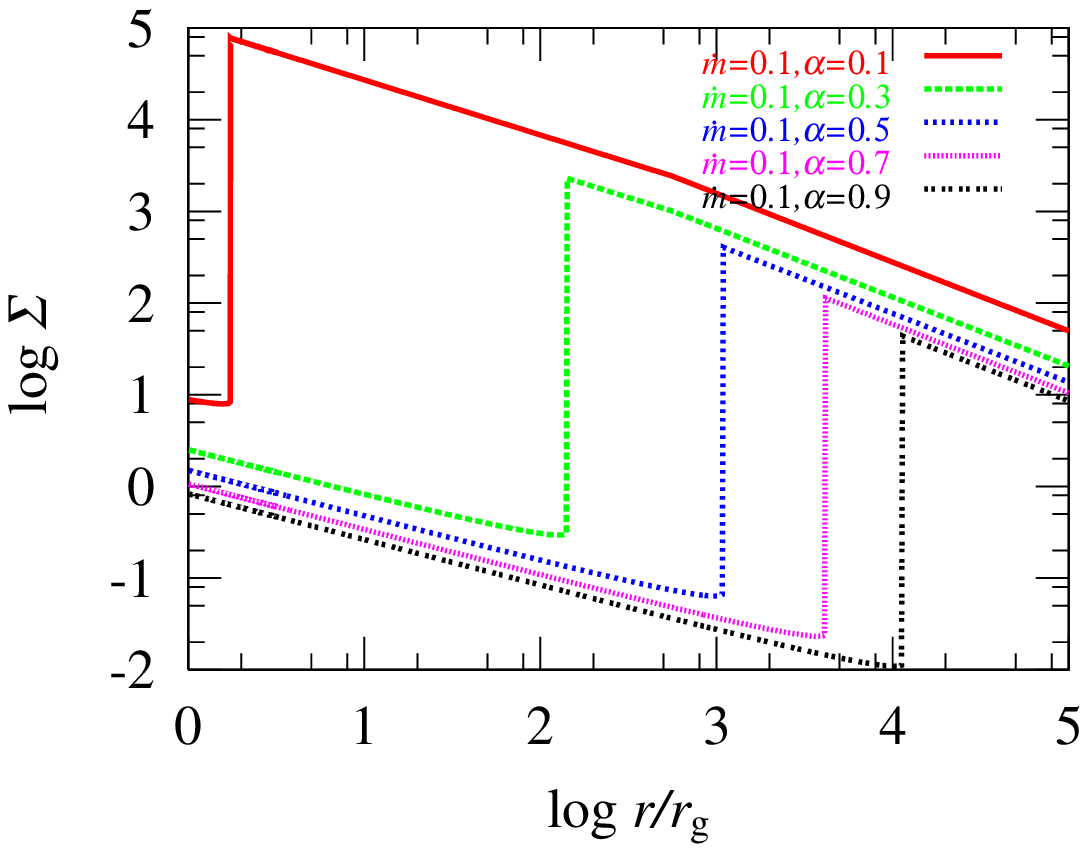}
    \FigureFile(60mm,60mm){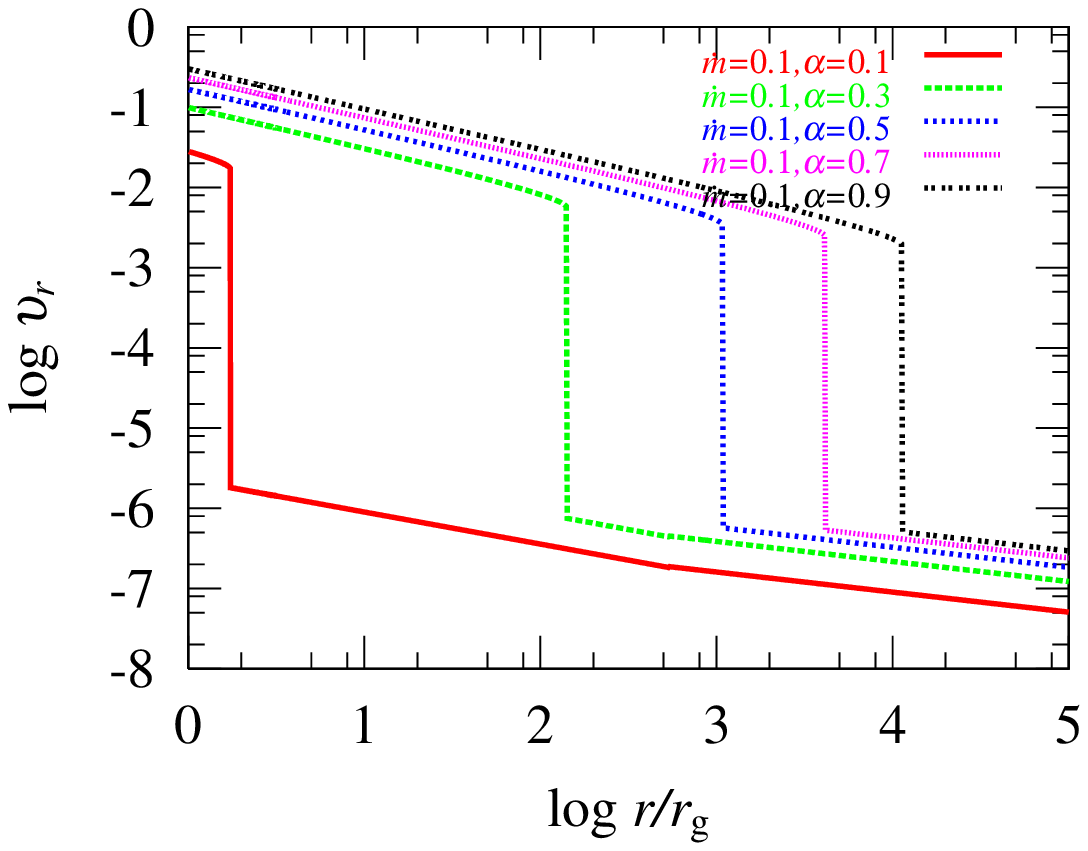}\\
    \FigureFile(60mm,60mm){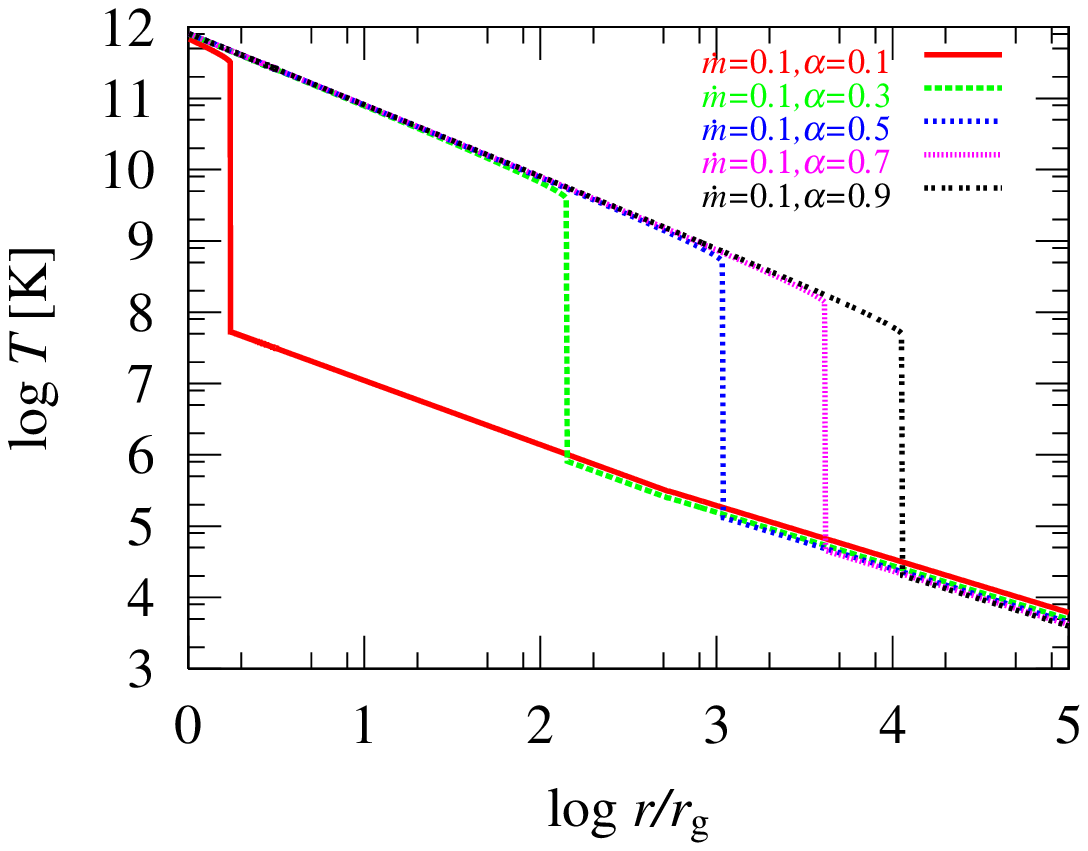}
    \FigureFile(60mm,60mm){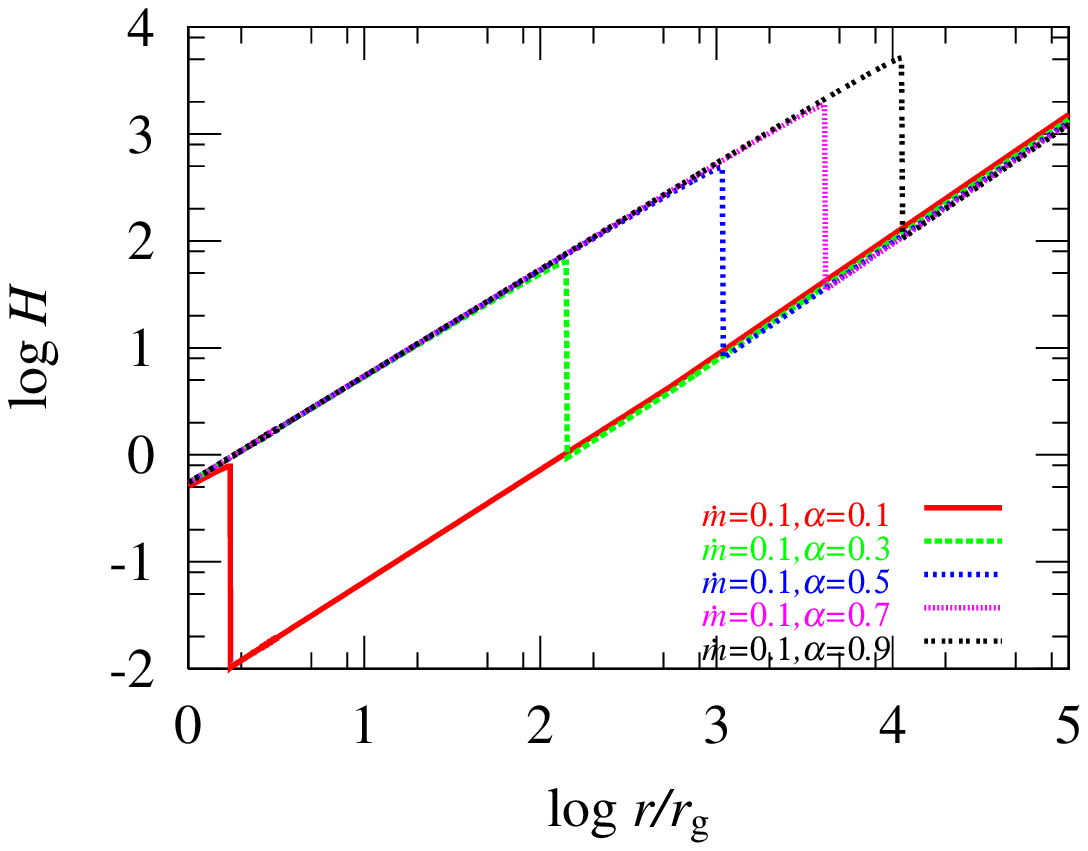}
  \end{center}
  \caption{Radial dependence of bimodal accretion flows with various $\alpha$.
  Panels are the surface density (top left), radial velocity (top right),
 gas temperature (bottom left), and scale height (bottom right), respectively. 
 We adopted the optically thick standard disk solutions in the disk outer region. }
\label{fig3}
\end{figure}

\subsection{Effect of Outer Boundary Conditions}
In deriving a solution for optically thin accretion flows,
 it is necessary to handle the initial value problem with care  
 because the outer boundary conditions can sometimes change the whole structure of the accretion flow.  
For example, a Keplerian angular frequency at the outer boundary
 was assumed in several early studies (Narayan et al. 1997)
 but Nakamura et al. (1996, 1997) assumed a sub-Keplerian angular momentum
 and a particular temperature at the outer boundary (see also Chen et al. 1997). 
In extreme cases, the outer boundary condition dramatically changes the entire disk structure,
 for example, when the outer boundary condition is set to high mach number and sub-Keplerian angular velocity
 (Yuan 1999; Yuan et al. 2000).
For this reason, it is not meaningful to compare the new analytical solutions with numerical solutions.

We suggest that the plausible outer boundary conditions must be determined from observations of
 the disk's outer regions in the objects being studied. 

\subsection{Stability of Transition Zone}

The time-dependent evolution of the transition zone, including the effects of turbulent energy
 transport, was recently computed by Gracia et al. (2003). 
They confirmed the result of Manmoto et al. (2000a), Manmoto \& Kato (2000b),
 and found that the transition zone is highly variable,
 suggesting that it may be the origin of the quasi-periodic oscillations  observed in Galactic X-ray binaries. 
However, if the magnetic field in the transition zone is taken into account, 
 the gas density is confined by the magnetic field, 
 which can cause the cooling energy at a given radius to change
 and potentially alter the location of the transition radius. 

Changes in the transition surface due to magnetorotational instability (MRI) 
 were examined by Kato (2000) using linear analysis. 
Linear MRI perturbations generated in the outer standard disk
 penetrate into the inner ADAF, which results in a highly variable transition zone. 
Recently, analytic solutions, including magnetic fields,
 have been investigated extensively by several authors 
(Meier 2005; Shadmehri \& Khajenabi 2005; Akizuki \& Fukue 2006). 
According to Machida et al. (2006), the magnetically dominant configuration survives for much longer than the thermal time, and is comparable to the accretion time scale, but it is shorter than the viscous time. However, such a long time calculation is still rare, so the result has not been reached consensus yet. We hope to be confirmed for a long time by the simulation a lot near the future.
 
It is possible that oscillation of the transition surface 
 produces the quasi-periodic oscillations commonly observed
 in Galactic black hole candidates (see also Kato 2000). 
This situation is similar to oscillations of the Alfv\'{e}n radius of the magnetosphere around a neutron star (Shapiro \& Teukolsky 1983). 
Thus MHD simulations of the region around the transition surface will be an intriguing topic 
 for further research. 

\section{Conclusion}

We described analytic formulae for accretion flows in the optically thin regime. 
The ratio of the advection cooling rate to the viscous heating rate is given by 
 the solution of a cubic equation, and is a function of $\hat{r}$, $\dot{m}$, and $\alpha$. 
Solving the cubic equation, we can obtain not only ADAF solutions,
 but also SLE-like solutions in a simple way. 
Our new solutions are physically valid,
 because both solutions originate from same cooling processes. 
The transition radius can be evaluated, where $f(\theta) = 0$,
 and is consistent with Honma's (1996) result. 
We expect that our analytic technique will be useful for order estimations
 of optically thin accretion flows; moreover,
 it will be useful to provide initial conditions for numerical simulations. 
 
\vspace{10mm}
We would like to thank an anonymous referee and S. Kato, and J. Fukue for helpful comments and discussions.
This work was supported in part by the Grants-in Aid of the
Ministry of Education, Science, Sports, and Culture of Japan
(16004706, KW).



\end{document}